\documentclass[aps,prl,superscriptaddress,longbibliography, showpacs,floatfix,letterpaper,preprintnumbers,nofootinbib,notitlepage,nobibnotes,reprint]{revtex4-2}
\usepackage{subcaption, caption}
\usepackage[colorlinks=true, linkcolor=blue, citecolor=blue, urlcolor=blue]{hyperref}
\usepackage{amssymb}
\usepackage{color}
\usepackage{graphicx} % Required for inserting images
\usepackage{bm}
\usepackage{amsmath,mathrsfs}
\usepackage{braket}
\usepackage{xfrac}
\usepackage{soul}
\usepackage{lineno}
\usepackage{slashed}
\usepackage{appendix}
\usepackage{tabularx,enumitem,paralist}
\usepackage{cancel} %cancels in colour
\usepackage{xcolor} %colours (\definecolor{name}{RGB}{R,G,B})

\newcommand{\polylog}[1]{\operatorname{Li}_{#1}}
\DeclareMathOperator{\arcsinh}{arcsinh}

\usepackage{float} % This package deals with placing figures where you want them, so that you don't have to "fight" with LaTeX.
\usepackage{orcidlink}

\begin{document}
\title{The high-energy behavior of tree-level scattering in finite-temperature QCD: estimates of theoretical systematic uncertainty in jet-medium Monte Carlo simulations}

\author{Lukas Opitz\,\orcidlink{0009-0008-0131-4307}}
\affiliation{Department of Physics, University of Regina, Regina, Saskatchewan S4S 0A2, Canada}

\author{Hemanth Regi\,\orcidlink{0009-0001-8564-159X}}
\affiliation{Department of Physics, University of Regina, Regina, Saskatchewan S4S 0A2, Canada}

\author{Gojko Vujanovic\,\orcidlink{0000-0001-5397-6662}}
\email[Corresponding author: ]{gojko.vujanovic@uregina.ca}
\affiliation{Department of Physics, University of Regina, Regina, Saskatchewan S4S 0A2, Canada}

\date{\today}

\begin{abstract}
We examine the behavior of tree-level scattering in thermal QCD at high energies and find significant deviations away from the commonly used approximations [\href{https://journals.aps.org/prd/abstract/10.1103/PhysRevD.44.1298}{Phys. Rev. D 44, 1298 (1991)},\href{https://journals.aps.org/prd/abstract/10.1103/PhysRevD.44.R2625}{Phys. Rev. D 44, R2625 (1991)}, \href{https://journals.aps.org/prd/abstract/10.1103/PhysRevD.77.014015}{Phys. Rev. D.77, 014015 (2008)}, \href{https://journals.aps.org/prd/abstract/10.1103/PhysRevD.77.114017}{Phys. Rev. D. 77.114017 (2008)}]. These deviations in the scattering rate also affect jet-medium transport coefficients at the partonic level, leading to a different kinematic dependence of the transverse momentum broadening per unit length $\hat{q}$. As scattering rates and $\hat{q}$ are used by large-scale Monte Carlo simulations of jets in the quark-gluon plasma (QGP), such as [\href{https://journals.aps.org/prc/10.1103/PhysRevC.111.054913}{Phys. Rev. C 111,054913 (2025)}], current constraints on $\hat q$ are biased owing to the approximations used therein. Besides theoretically improving $\hat q$, the differences between the $\hat q$ herein and the approximate $\hat q$ used in jet Monte Carlo simulations are used to construct a theoretical systematic uncertainty, which in turn can be employed to devise a covariance matrix for $\hat q$ and enables updating the uncertainty bands on $\hat q$ obtained by Bayesian analysis.
\end{abstract}

\maketitle

\textbf{\textit{Introduction.}}---Relativistic heavy-ion collisions that produce a quark-gluon plasma (QGP) have been the best environment to test relativistic quantum field theories at finite temperature, more specifically of QCD. Beside the QED plasma, the QGP is the only plasma of fundamental (i.e. SU(3)) interactions that can be created in the laboratory. Creating an electroweak plasma is currently outside of our experimental capabilities. Detailed comparisons of quantum field theory and effective field theory simulations against data of tomographic probes of relativistic heavy-ion collisions, such as dilepton production~\cite{Wu:2024vyc,Churchill:2023zkk,Churchill:2023vpt,Vujanovic:2019yih} and jet quenching \cite{JETSCAPE:2024cqe,JETSCAPE:2023ikg,JETSCAPE:2022jer,JETSCAPE:2021ehl}, furthers our understanding of field theories in thermal equilibrium and beyond by providing ways to validate theoretical calculations. This contribution focuses on furthering our understanding of jet-medium interactions. 

Understanding the quenching of QCD jets inside of the QGP is now entering the era of precision exploration using Bayesian analyses \cite{JETSCAPE:2024cqe,Kumar:2025asj,Kumar:2025egh,Sirimanna:2022zje,Modarresi-Yazdi:2024vfh,Mehtar-Tani:2022zwf}. Depending on the kinematic regime, different models are used, thus maximizing simulation accuracy. Quantifying the accuracy of simulations employed in Bayesian model-to-data comparisons is thus crucial, specifically a better control of theoretical systematic uncertainties present within model calculations is needed. Accounting for theoretical systematic uncertainties increases the reliability of the constraints obtained on jet-medium transport coefficients, dissipative processes in the QGP and so on. Bayesian constraints on dissipative processes in the QGP bulk, such as its specific shear and bulk viscosity, have recently been obtained for different prescriptions for converting fluid fields into particles \cite{JETSCAPE:2020shq}, while model discrepancy in Bayesian emulators \cite{Jaiswal:2025deb,Jaiswal:2025hyp} allows to further incorporate theoretical systematic uncertainty, in situations where a first principles evaluation of theoretical systematic uncertainty are difficult to ascertain.

To date, a quantitative and systematic accounting of theoretical uncertainties associated with tree-level jet-medium interactions and their transport coefficients in the QGP has not been done, be it using Bayesian model averaging \cite{JETSCAPE:2020shq}, or by model discrepancies \cite{Jaiswal:2025deb,Jaiswal:2025hyp}. Instead, qualitative comparisons between various theoretical calculations are used (see Ref.~\cite{Cao:2024pxc} for a recent comparison of jet-medium calculations). A first step towards fully addressing this discrepancy is presented herein. 

An important transport coefficient quantifying jet-medium interactions is $\hat{q}$, accounting for medium-induced transverse momentum broadening of jets in the QGP. Specifically, $\hat q=\langle q^2_\perp\rangle_L/L$ is a measure of the average transverse momentum squared $q^2_\perp$ exchanged between particles in the jet and those in the QGP per unit length \cite{Majumder:2009ge,Caron-Huot:2009fku,Caron-Huot:2008zna}. The transverse direction is chosen to be orthogonal to the jet direction. In many numerical simulations of jets in the QGP, $\hat{q}$ is calculated as the second moment of the $a+b\to 1+2$ scattering rate between the jet $(a,1)$ and the QGP $(b,2)$ through the exchange of a gluon, where~\footnote{$\hat q$ in Eq.~(\ref{eq:qhat_matrix}) is $(2\pi)^3$ times smaller than the one in Ref.~\cite{JETSCAPE:2024cqe}, stemming from $\frac{d^4 q}{(2\pi)^3}$ being used herein.}
\begin{align}
\hat{q}&= \frac{1}{2E_a}\left[\prod_{i=b,1,2}\int\frac{d^3 p_i}{\left(2\pi\right)^3 2E_i}\right] \int \frac{d^4 q}{(2\pi)^3} f_b\left(p_b\right) \left[1\pm f_2\left(p_2\right)\right]\nonumber\\
&\times(2\pi)^4\delta^{(4)}\left(p_a-p_1-q\right)\delta^{(4)}\left(p_b+q-p_2\right) q^2_\perp \overline{\left|\mathcal{M}_{a,b\to1,2}\right|^2}\nonumber\\
&=\frac{\langle q^2_\perp \rangle_L}{L}.
\label{eq:qhat_matrix}
\end{align}
In Eq.~(\ref{eq:qhat_matrix}), $f_b(p_b)$ and $f_2(p_2)$ correspond to the distribution of particles in the QGP, before and after scattering, respectively. In thermal equilibrium, which is the case considered here, $f_{b,2}$ is either the Bose-Einstein or the Fermi-Dirac distribution, depending on the type of particle, while Bose-Enhancement and Pauli-Blocking were accounted for via the $1\pm$ term preceding $f_2$. Finally $\overline{\left|\mathcal{M}_{a,b\to1,2}\right|^2}$ is the scattering matrix element. The focus herein is on improving the high energy (i.e. ultraviolet) behavior of $2\to2$ scattering by going well beyond the leading logarithm (leading-log for short) approximation. In the high-energy limit, infrared effects, that are captured by Hard Thermal Loop (HTL) resummations, play less of a role and will be considered in an upcoming publication. 

On the theoretical side, $\hat{q}$ for $gg\to gg$ scattering of the form
\begin{align}
\frac{\hat{q}}{T^3}&=\frac{9\alpha^{2}_s\zeta(3)}{\pi^4}\ln\left[\frac{2E_a T}{m^2_D}\right]+O\left(e^{-E_a/T}\right),
\label{eq:qhat_leading-log}
\end{align}
was considered in \cite{JETSCAPE:2024cqe}, where $E_a$ is the incoming jet parton energy, $m_D=\sqrt{6\pi\alpha_s}\,T$ is the Debye screening mass for a three-flavor QGP at temperature $T$, and $\zeta(3)\approx1.2020569031$ is Ap\'ery's constant \cite{Cohl:2014drm}. Our goal is to improve the behavior of the scattering rate, $\hat{q}$, and the longitudinal energy loss $\hat e$ by going beyond the leading-log (eikonal) energy dependence $\hat q\sim\ln\left[\frac{2E_a T}{m^2_D}\right]$, thus improving jet-medium Monte Carlo simulations. Using the full tree-level result, theoretical systematic uncertainty will be constructed through the difference between the full tree-level and any approximate result, and thus can be included as a theory uncertainty via a covariance matrix in future Bayesian analysis. Without HTL resummations, the leading-log behavior becomes $\sim\ln\left[\frac{q^{\rm max}_\perp}{q_c}\right]$ \cite{Caron-Huot:2009fku}, where $q^{\rm max}_\perp$ is some maximum exchanged momentum, while $q_c$ is some cut-off momentum, below which HTL resummations \cite{Pisarski:1989cs, Braaten:1989kk, Braaten:1989kk, Braaten:1989mz, Braaten:1990az} are necessary. In this paper, the Debye mass is playing the role of the cut-off momentum, i.e. $q_c\equiv m_D$, which avoids the needed inclusion of many-body effects of HTL that screen the divergences in the tree-level $2\to2$ matrix element. These cut-off effects will be accounted for in an upcoming study that revisits combining HTL with tree-level results. In the high-energy limit, the leading-log energy loss has been computed by Braaten and Thoma \cite{Braaten:1991jj,Braaten:1991we}, while the first subeikonal (beyond leading-log) correction was devised by Peign\'e and Peshier \cite{Peigne:2007sd,Peigne:2008nd}. In the infrared limit, the leading order HTL \cite{Peigne:2008nd} results were enhanced by next-to-leading order corrections \cite{Caron-Huot:2008zna,Caron-Huot:2009fku,Ghiglieri:2015zma,Ghiglieri:2015ala}. 

To illustrate the change in the scattering rate, $\hat{q}$, and $\hat e$ obtained by considering the full tree-level matrix elements and kinematic phase spaces outside of HTL, the $gg\to gg$ channel is investigated. The kinds of changes found herein persist in other $2\to2$ processes. Particular focus is given towards identifying power suppressed corrections to the rate, $\hat{q}$, and $\hat e$ beyond the $1/E$ correction of Refs.~\cite{Peigne:2007sd,Peigne:2008nd}. Thus, exponentially suppressed terms -- $[1+f_2]$ in Eq.~(\ref{eq:qhat_matrix}) -- from the Bose-enhancement in the $gg\to gg$ channel are neglected in our analytical result, as they are parametrically much smaller than ones behaving as $E^{-n}$ ($n\in \mathbb{N})$. Bose-enhancement effects are considered numerically, however, with their analytical analysis being revisited soon~\cite{long_paper_in_prep}. As the jet shower develops and partons in the jet start exploring a larger phase space, $E^{-n}$ suppressed terms will contribute to the scattering and radiation probability used within multi-scale jet Monte-Carlo simulation, and therefore should be included in such simulations as much as possible.

\textbf{\textit{Scattering rate.}}---The full tree-level scattering rate is given by
\begin{align}
\frac{d^3R}{d^3p_a}&=\frac{1}{16(2\pi)^7E_a^2}\int^{\infty}_0 dE_b\int^{E_a+E_b}_0 dE_2\nonumber\\\
&\times\int^{0}_{t_\mathrm{min}} dt\int^{s_+}_{s_-} ds \frac{f_b[1\pm f_2]\overline{\left|\mathcal{M}_{a,b\to1,2}\right|^2}(s,t)}{\sqrt{\Lambda(s,t)}},\nonumber\\
\Lambda(s,t)&=st(s+t)-\left(E_b-E_2\right)^2s^2-\left(E_a+E_b\right)^2t^2\nonumber\\
&-2\left[E_aE_2+E_bE_1\right]st,\\
s_{\pm}&=\frac{t[t-2(E_aE_2+E_bE_1)]}{\left(E_b-E_2\right)^2-t}\nonumber\\&\mp\frac{\sqrt{\left[t+4E_aE_1\right]\left[t+4E_bE_1\right]}}{\left(E_b-E_2\right)^2-t},\nonumber\\
t_\mathrm{min}&=\left(E_b-E_2\right)^2-\min\left[\left(E_a+E_1\right)^2,\left(E_b+E_2\right)^2\right],\nonumber
\end{align}
where $s$, $t$, and $u$ are the usual Mandelstam variables, and $E_1=E_a+E_b-E_2$. The matrix element $\overline{\lvert\mathcal{M}_{gg\to gg}\rvert^2}(s,t)=288\alpha_s^2\left(2\pi\right)^2\left(3-\frac{tu}{s^2}-\frac{su}{t^2}-\frac{st}{u^2}\right)$, has a pole at $t=0$ and $u=0$, for a non-vanishing $p_a$. To regulate this pole, we change the range of $t$ such that $t\in[t_\mathrm{min},0]\to t\in[t_\mathrm{min},-m^2_D]$, where $m_D$ is a cut-off set to the Debye mass, and excises this pole. The $s$ and $t$ integrals can thus be performed exactly, yielding
\begin{align}
\frac{d^3R}{d^3p_a}&=\frac{9\alpha_s^2}{4\pi^4E_a^2}\int_0^\infty dE_b\frac{1}{e^{\beta E_b}-1}\\
&\times\int_0^{E_a+E_b} dE_2\frac{1}{1-e^{-\beta E_2}}\left[\mathbf{I}+\mathbf{II}+\mathbf{III}+\mathbf{IV}\right],\nonumber
\end{align}
where
\begin{align}
\mathbf I&=3\min[E_a,E_b,E_a+E_b-E_2,E_2],\nonumber\\
\mathbf{II}&=-\frac{E_aE_b\left(E_a+E_b-E_2\right)E_2}{\left(E_a+E_b\right)^3},\nonumber\\
\mathbf{III}&=\frac{E_aE_b\left(E_a+E_b-E_2\right)E_2}{\left[\left(E_b-E_2\right)^2+m_D^2\right]^\frac{3}{2}},\nonumber\\
\mathbf{IV}&=\frac{E_aE_b\left(E_a+E_b-E_2\right)E_2}{\left[\left(E_a-E_2\right)^2+m_D^2\right]^\frac{3}{2}},
\end{align}
while $\beta=1/T$. The four-point gluon vertex, the $s$-channel, $t$-channel, and $u$-channel diagrams correspond to $\mathbf{I},\mathbf{II},\mathbf{III}$, and $\mathbf{IV}$, respectively, thus matching the corresponding terms in $\overline{\lvert\mathcal{M}_{gg\to gg}\rvert^2}$. 

In the high-energy $\beta E_a>1$ limit, only the the $t$- and $u$-channel contribute, and $\left[1-e^{-\beta E_2}\right]^{-1}\approx1+O\left(e^{-\beta E_2}\right)$ approximates the exact scattering rate well. The contributions from the other two channels are considered in~\cite{long_paper_in_prep}. To be consistent with Eq.~(\ref{eq:qhat_leading-log}), exponentially suppressed terms are neglected~\footnote{While Eq.~(\ref{eq:qhat_leading-log}) takes the limit $[1+f_2]\to 1$, Bose-enhancement affects the result via $\zeta(3)\to\zeta(2)$.}, thus giving  
\begin{align}
\frac{d^3R_{t+u}}{d^3 p_a} \simeq \frac{9\alpha^2_s T}{\pi^4}&\left\{\frac{A^-_1(z)}{2z^2}+\frac{\zeta(3)}{ z^2}\frac{\sqrt{x^2_a+z^2}}{x_a}-\frac{B^-_1(z)}{2x_a}\right.\nonumber\\
&\quad\left.-\frac{\pi^2}{12x_a}\arcsinh\left(\frac{x_a}{z}\right)\right\}+O\left(e^{-x_2}\right)
\label{eq:rate}\\
\alpha^{-}_n\left(c,\epsilon;x_a,z\right)&\equiv\int^c_\epsilon dx_b\frac{x^n_b\sqrt{\left(x_a-x_b\right)^2+z^2}}{e^{x_b}-1}\quad\forall n\in\mathbb{N}_0,\nonumber\\
A^{-}_n\left(\epsilon;z\right)&=\alpha^-_n(\infty,\epsilon;0,z)\equiv\lim_{c\to\infty}[\alpha^{-}_n(c,\epsilon;0,z)] \nonumber\\
A^{-}_n\left(z\right)&=\lim_{\epsilon\to0^+}\left[A^-_n(\epsilon;z)\right]\nonumber\\
\beta^{-}_n(c,\epsilon;x_a,z)&\equiv\int^c_\epsilon dx_b\frac{x^n_b\,\,\arcsinh\left(\frac{x_a-x_b}{z}\right)}{e^{x_b}-1}\quad\forall n\in\mathbb{N}_0,\nonumber\\
-B^{-}_n(\epsilon;z)&=\beta^{-}_n(\infty,\epsilon;0,z)\equiv\lim_{c\to\infty}[\beta^-_n(c,\epsilon;0,z)]\nonumber\\
-B^{-}_n(z)&=\lim_{\epsilon\to0^+}[\beta^{-}_n(\infty,\epsilon;0,z)]
\label{eq:tree-level_asymptotics}
\end{align}
where $\epsilon>0$, $x_a=E_a/T$, $x_b=E_b/T$, $x_2=E_2/T$ and $z=m_D/T$. The minus superscript atop $\alpha^-_n,\beta^-_n,A^-_n,B^-_n$ is there to remind the reader of the Bose-Einstein thermal distribution at play. To obtain $A^-_1(z)$ and $B^-_1(z)$ in the ultraviolet limit, one uses 
\begin{align}
\arcsinh\left(\frac{x}{z}\right)&\simeq \ln\left(\frac{x}{z}\right)+\ln(2)+\frac{z^2}{2x^2}+O\left(\frac{z^4}{x^{4}}\right)\nonumber\\
\sqrt{1+\frac{z^2}{x^{2}}}&\simeq 1+\frac{z^2}{2x^2}+O\left(\frac{z^4}{x^{4}}\right)
\label{eq:approx}
\end{align}
yielding
\begin{align}
A^-_1(z)&\simeq \int^\infty_0 dx_b \frac{x^2_b}{e^{x_b}-1}+O\left(\frac{z^2}{x^{2}_a}\right)= 2\zeta(3)+O\left(\frac{z^2}{x^{2}_a}\right)\nonumber\\
B^-_1(z)&\simeq \int^\infty_0 dx_b \frac{x_b\left[\ln(\frac{x_b}{z})+\ln(2)\right]}{e^{x_b}-1}+O\left(\frac{z^2}{x^{2}_a}\right)\nonumber\\
&\approxeq\frac{\pi^2}{6}\left[\ln\left(\frac{2}{z}\right)+1-\gamma\right]+\zeta'(2)+O\left(\frac{z^2}{x^{2}_a}\right)
\label{eq:A_1_B_1_approx}
\end{align}
where $\gamma\approx0.5772156649$ is the Euler-Mascheroni constant, while the derivative of the Riemann zeta function $\zeta'$ gives $\zeta'(2)=\frac{\pi^2}{6}\left[\gamma+\ln\left(2\pi\right) -12\ln\left(A\right)\right]$, with $A\approx 1.282427129$ being the Glaisher–Kinkelin constant \cite{Cohl:2014drm}.

\begin{figure}[H]
	\centering
	\includegraphics[width=\linewidth]{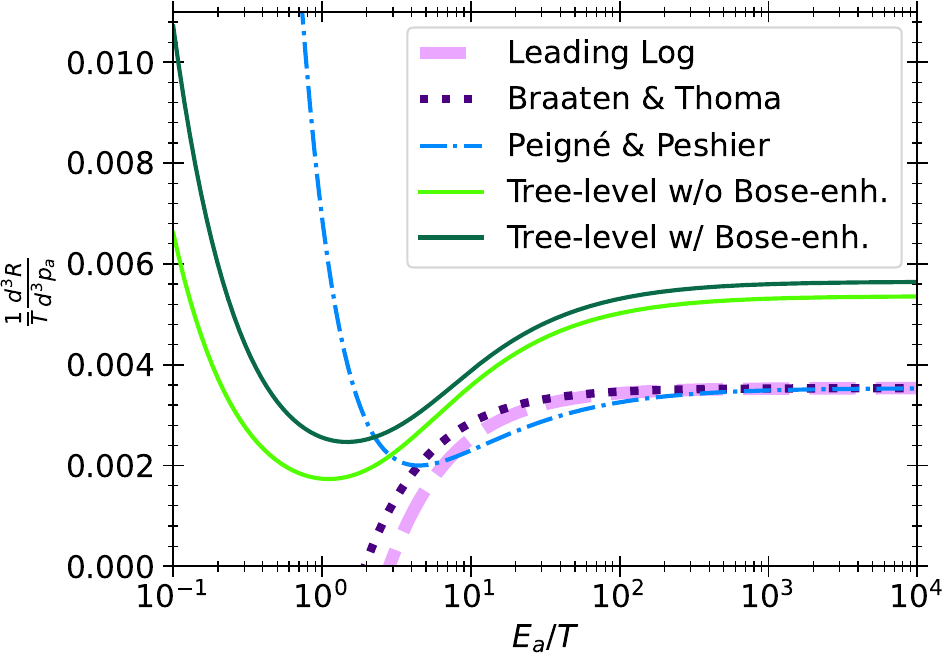}
    \label{fig:rate_x}
    \caption{Scattering rate for $gg\to gg$ process at $\alpha_s=0.3$.}
\label{fig:rate}
\end{figure}

The ultraviolet behavior of the leading-log approximation, where $q^2\approx -q^2_\perp$, while $E_b\gg q_\perp$ and $E_a>E_b$ \cite{long_paper_in_prep}, gives
\begin{align}
\frac{d^3R_{t+u}}{d^3 p_a}&\simeq \frac{9\alpha_s^2T}{\pi^4}\left[\frac{2\zeta(3)}{z^2}-\frac{\zeta(3)}{x_a}\right]+O\left(e^{-x_a}\right).%-e^{-x_a}\left[\frac{x^2_a}{z^2}+\frac{2x_a}{z^2}\right.\right.\nonumber\\
%&+\left.\left.\frac{2}{z^2}-1-\frac{2}{x_a}-\frac{2}{x^2_a}\right],
%\frac{18\alpha_s^2T\zeta(3)}{\pi^4 z^2}
\label{eq:rate_LL}
\end{align}
The Braaten \& Thoma \cite{Braaten:1991jj,Braaten:1991we} approach gives
\begin{align}
\frac{d^3R_{t+u}}{d^3 p_a}&\simeq\frac{9\alpha_s^2T}{\pi^4}\left[\frac{2\zeta(3)}{z^2}-\frac{\pi^2}{12x_a}\right].
\label{eq:rate_BT}
\end{align}
Both Eq.~(\ref{eq:rate_LL}) and (\ref{eq:rate_BT}) are missing $x^{-1}_a\arcsinh(x_a/z)$ dependence. More mathematical details on these approximations are presented in Ref.~\cite{long_paper_in_prep}. The first correction to the Braaten \& Thoma's result was formally devised by Peign\'e \& Peshier \cite{Peigne:2007sd,Peigne:2008nd}, which, for the gluon-gluon scattering channel, gives
\begin{align}
\frac{d^3R_{t+u}}{d^3p_a} \simeq\frac{9\alpha_s^2T}{\pi^4}&\left[\frac{2\zeta(3)}{z^2}-\frac{\pi^2}{12x_a}\ln\left(\frac{4x_a}{z^2}\right)-\frac{\zeta'(2)}{2x_a}\right.\nonumber\\
&\left.-\frac{(1-\gamma)\pi^2}{12x_a}\right]+O\left(z^2x^{-2}_a\right).
\label{eq:rate_PP}
\end{align}
Comparing Eq.~(\ref{eq:rate_PP}) with Eq.~(\ref{eq:rate}), one can see how approximating Eq.~(\ref{eq:rate}) using Eq.~(\ref{eq:A_1_B_1_approx}) leads to Eq.~(\ref{eq:rate_PP}). It is apparent that the approach in Refs.~\cite{Peigne:2007sd,Peigne:2008nd} misses the $x^{-1}_a\sqrt{x^2_a+z^2}$ term present in Eq.~(\ref{eq:rate}) and only partially captures the $x^{-1}_a \arcsinh(x_a/z)$ via $x^{-1}_a\ln(x_a/z)$. At intermediate energies, such differences likely affect the scattering probability of gluons in the jet shower as well as their medium-modified (and medium-induced) radiation spectrum, affecting jet substructure, and likely biasing current jet-medium Bayesian analysis. To address this issue, Bayesian model-to-data comparisons will need to be revised by accounting for the rates provided herein. The scattering rate is plotted in Fig.~\ref{fig:rate} as a function of $x_a$.  
The difference in the ultraviolet asymptotic value of the scattering rate, when comparing the Peign\'e \& Peshier result in Eq.~(\ref{eq:rate_PP}) and ours in Eq.~(\ref{eq:rate}), stems from resumming $O\left(z^2\right)$ corrections within $A^-_1(z)$, resulting in large corrections seen in Fig.~\ref{fig:rate}. For the sake of completeness, Bose-enhancement effects on the scattering rate are also shown in  Fig.~\ref{fig:rate}. When comparing the two solid lines, one sees that Bose-enhancement increases the scattering rate by $\sim 5-10$\% for $10\lesssim x_a\lesssim100$ with a more modest increase of $\sim 5$\% for $x_a\gtrsim 100$.

\textbf{\textit{Transverse momentum broadening $\hat{q}$.}}--- In the ultraviolet limit, $t$- and $u$-channel contributions to $\overline{\lvert\mathcal{M}_{gg\to gg}\rvert^2}$ produce $\langle q^2_\perp\rangle_L/L$, which are labeled as $\hat q_t$ and $\hat q_u$ below, and dominate over all other channels. The $\hat q$ in Eq.~(\ref{eq:qhat_leading-log}), when used in the higher-twist formalism \cite{Guo:1998rd,Majumder:2009ge,Kumar:2025asj,Kumar:2025egh}, corresponds to solely considering $\hat q_t$ here, and
\begin{widetext}
\begin{align}
\hat{q}_t&\simeq\frac{9\alpha^2_s T^3}{2\pi^4}\left\{\left[\frac{\pi^2}{12}-\frac{11\zeta(3)}{3x_a}-\frac{13\pi^2z^2}{24x^2_a}+\frac{4\zeta(3)z^2}{3x^3_a}\right]\sqrt{x^2_a+z^2}+\left[2\zeta(3)+\frac{\pi^2z^2}{2x_a}-\frac{3\zeta(3)z^2}{x^2_a}-\frac{\pi^2 z^4}{8x^3_a}\right]\arcsinh\left(\frac{x_a}{z}\right)-\zeta(3)\right.\nonumber\\
&\left.-\frac{\pi^4}{18x_a}+B^-_2(z)+\frac{3z^2}{x_a}B^-_1(z)-\frac{3z^2}{2x^2_a}B^-_2(z)-\frac{3z^4}{4x^3_a}B^-_1(z)-\frac{3A^-_1(z)}{2}+\frac{A^-_2(z)}{x_a}+\frac{A^-_3(z)}{4x^2_a}+\frac{23z^2}{8x^2_a}A^-_1(z)+\frac{A^-_4(z)}{30x^3_a}-\frac{11z^2}{60x^3_a}A^-_2(z)\right.\nonumber\\
&\left.-\left[\frac{z^2}{2}-\frac{9z^4}{8x^2_a}\right]B^-_0(z)+z\left[2g^-_1(x_a)-\frac{2g^-_2(x_a)}{x_a}-\frac{4z^2}{x^2_a}g^-_1(x_a)+\frac{2z^2}{3x^3_a}g^-_2(x_a)\right]-z^2\left[\frac{3}{x_a}\Delta\beta^-_1-\frac{3}{2x^2_a}\Delta\beta^-_2-\frac{3z^2}{4x^3_a}\Delta\beta^-_1\right]\right.\nonumber\\
&\left.-\frac{11}{20}\Delta\alpha^-_1+\frac{17}{60x_a}\Delta\alpha^-_2+\frac{17}{60x^2_a}\Delta\alpha^-_3+\frac{319z^2}{120x^2_a}\Delta\alpha^-_1+\frac{\Delta\alpha^-_4}{30x^3_a}-\frac{11z^2}{60x^3_a}\Delta\alpha^-_2+\left[\frac{z^2}{2}-\frac{9z^2}{8x^2_a}\right]\lim_{\epsilon\to0^+}\left[\Delta\beta^-_0(\epsilon)+\arcsinh\left(\frac{x_a}{z}\right)\ln(\epsilon)\right]\right.\nonumber\\
&\left.-\left[\frac{2z^2}{x_a}-\frac{8z^4}{15x^3_a}\right]\left[\lim_{\epsilon\to0^+}\left[z\ln\left(\epsilon\right)+A^-_0\left(\epsilon;z\right)\right]-2z\ln\left(1-e^{-x_a}\right)\right]-\left[\frac{x_a}{20}+\frac{137z^2}{120x_a}-\frac{8z^4}{15x^3_a}\right]\lim_{\epsilon\to0^+}\left[\Delta\alpha^-_0(\epsilon)+\sqrt{x^2_a+z^2}\ln\left(\epsilon\right)\right]\right\}\nonumber\\
&+\,\,O\left(e^{-x_2}\right),\label{eq:qhat_t}\\
\hat{q}_u&\simeq\frac{9\alpha_s^2T^3}{2\pi^4}\left\{\left[\frac{8\zeta(5)}{z^2x_a}-\frac{5\zeta(3)}{x_a}-\frac{\pi^4}{30x^2_a}-\frac{4\zeta(5)}{x^3_a}\right]\sqrt{x^2_a+z^2}+\left[2\zeta(3)-\frac{\pi^4}{15x_a}-\frac{3\zeta(3)z^2}{x^2_a}-\frac{\pi^4z^2}{10x^3_a}-\frac{12\zeta(5)z^2}{x^4_a}\right]\arcsinh\left(\frac{x_a}{z}\right)\right.\nonumber\\
&\left.+\frac{A^-_3(z)}{3z^2}-\frac{2}{3}A^-_1(z)+\frac{3}{2x_a}A^-_2(z)+z\left[\frac{\pi^2}{9}+\frac{4\zeta(3)}{x_a}+\frac{2\pi^4}{15x^2_a}+\frac{16\zeta(5)}{x^3_a}\right]+B^-_2(z)-\frac{B^-_3(z)}{x_a}+\frac{z^2B^-_1(z)}{2x_a}\right\}+O\left(e^{-x_2}\right)
\label{eq:qhat_u}
\end{align}
\end{widetext}
where
\begin{align}
\Delta\alpha^-_n(\epsilon)&=2\alpha^-_n\left(x_a,\epsilon;x_a,z\right)-\alpha^-_n\left(\infty,\epsilon;x_a,z\right),\nonumber\\\
\Delta\alpha^-_n&=\lim_{\epsilon\to 0^+}\left[\Delta\alpha^-_n\left(\epsilon\right)\right],\nonumber\\
\Delta\beta^-_n(\epsilon)&=2\beta^-_n\left(x_a,\epsilon;x_a,z\right)-\beta^-_n\left(\infty,\epsilon;x_a,z\right),\nonumber\\
\Delta\beta^-_n&=\lim_{\epsilon\to 0^+}\left[\Delta\beta^-_n\left(\epsilon\right)\right],\nonumber\\
g^-_1\left(x_a\right)&=2x_a\ln\left(1-e^{-x_a}\right)-2\polylog{2}\left(e^{-x_a}\right)+\frac{\pi^2}{6},\nonumber\\
g^-_2(x_a)&=2\zeta(3)+2x^2_a\ln\left(1-e^{-x_a}\right)-4x_a\polylog{2}\left(e^{-x_a}\right)\nonumber\\
&-4\polylog{3}\left(e^{-x_a}\right),
\end{align}
and $\polylog{n}$ is the polylogarithm given by \cite{Cohl:2014drm} 
\begin{align}
\polylog{n}\left(e^{x_a}\right)&=\frac{1}{\Gamma(n)}\int^\infty_0 dt \frac{t^{n-1}}{e^{t-x_a}-1} \,\,\,\forall\,\, {\rm Re}(n)>0,\nonumber\\
\Gamma(n)&=\int^\infty_0 dt\, t^{n-1} \,e^{-t}\,\,\,\forall\,\, {\rm Re}(n)>0.
\end{align}
Note that $\alpha^-_n$ and $\beta^-_n$ are defined in Eq.~(\ref{eq:tree-level_asymptotics}). 

Using the leading-log approximation \cite{long_paper_in_prep}, $q^2\approx -q^2_\perp$, $E_b\gg q_\perp$ and $E_a>E_b$, is giving $\hat q_t$ in Eq.~(\ref{eq:qhat_leading-log}), while $\hat q_u=\hat q_t$. The Braaten \& Thoma approach yields 
\begin{align}
\hat{q}_t&\simeq\frac{9\alpha_s^2T^3}{2\pi^4}\left[2\zeta(3)\ln\left(\frac{2x_a}{z^2}\right)+2\zeta'(3)+2\left[\ln(2)-\gamma\right]\zeta(3)\right.\nonumber\\
&\left.\qquad\qquad\quad+\frac{\pi^4}{45x_a}\right]+O\left(\frac{z^2}{x^{2}_a}\right),\\
\hat{q}_u&\simeq\frac{9\alpha_s^2T^3}{2\pi^4}\left\{\left[2\zeta(3)-\frac{2\pi^4}{45x_a}\right]\ln\left(\frac{2x_a}{z^2}\right)+2\zeta'(3)\right.\nonumber\\
&\left.\qquad\qquad\quad+2\left[\ln (2)-\gamma\right]\zeta(3)+\frac{16\zeta(5)}{z^2}-\frac{4\zeta'(4)}{x_a}\right.\nonumber\\
&\left.\qquad\qquad\quad+\left[\gamma-1-\ln(2)\right]\frac{2\pi^4}{45x_a}\right\}+O\left(\frac{z^2}{x^{2}_a}\right).\nonumber\\
\end{align}
The Peign\'e \& Peshier approach generates
\begin{align}
\hat{q}_t&\simeq\frac{9\alpha_s^2T^3}{2\pi^4}\bigg[2\zeta(3)\ln\left(\frac{2x_a}{z^2}\right)+2\zeta'(3)+\frac{\pi^4}{90x_a}\label{eq:qhat_t_PP}\nonumber\\
&\qquad\qquad\quad+\left[2\ln(2)-2\gamma-1\right]\zeta(3)\bigg]+O\left(\frac{z^2}{x^{2}_a}\right),\nonumber\\
\end{align}
\begin{align}
\hat{q}_u&\simeq\frac{9\alpha_s^2T^3}{2\pi^4}\left\{\left[2\zeta(3)-\frac{\pi^4}{15x_a}\right]\ln\left(\frac{2x_a}{z^2}\right)+2\zeta'(3)\right.\nonumber\\
&\left.\qquad\qquad+\left[2\ln(2)-2\gamma-1\right]\zeta(3)+\frac{16\zeta(5)}{z^2}-\frac{6\zeta'(4)}{x_a}\right.\nonumber\\
&\left.\qquad\qquad+\left[2\gamma-1-2\ln(2)\right]\frac{\pi^4}{30x_a}\right\}+O\left(\frac{z^2}{x^2_a}\right)\label{eq:qhat_u_PP}
\end{align}
The expression in Eq.~(\ref{eq:qhat_t_PP}) still lacks more subtle dependence in $\hat q_t$ present in Eq.~(\ref{eq:qhat_t}). To express deviations away from the leading-log result, the variable $\Delta \hat q$ is used.~\footnote{Differences between Braaten and Thoma and Peign\'e and Peshier approaches result in distinct $O(1)$ coefficients, as well as coefficients in front of $x^{-1}_a$ terms and $z^{-2}$ terms --- more details are in Ref.~\cite{long_paper_in_prep}.} It is defined as 
\begin{align}
\Delta\hat{q}\equiv\frac{|\hat{q}-\hat{q}_\text{LL}|}{|\hat{q}_\text{LL}|}
\label{eq:Delta_qhat}
\end{align}
where $\hat q$ refers to any of the approaches used, from Eq.~(\ref{eq:qhat_t}) to Eq.~(\ref{eq:qhat_t_PP}), while $\hat{q}_{\rm LL}$ refers to Eq.~(\ref{eq:qhat_leading-log}). 

\begin{figure}[H]
\centering
\begin{subfigure}{\linewidth}
	\includegraphics[width=\linewidth]{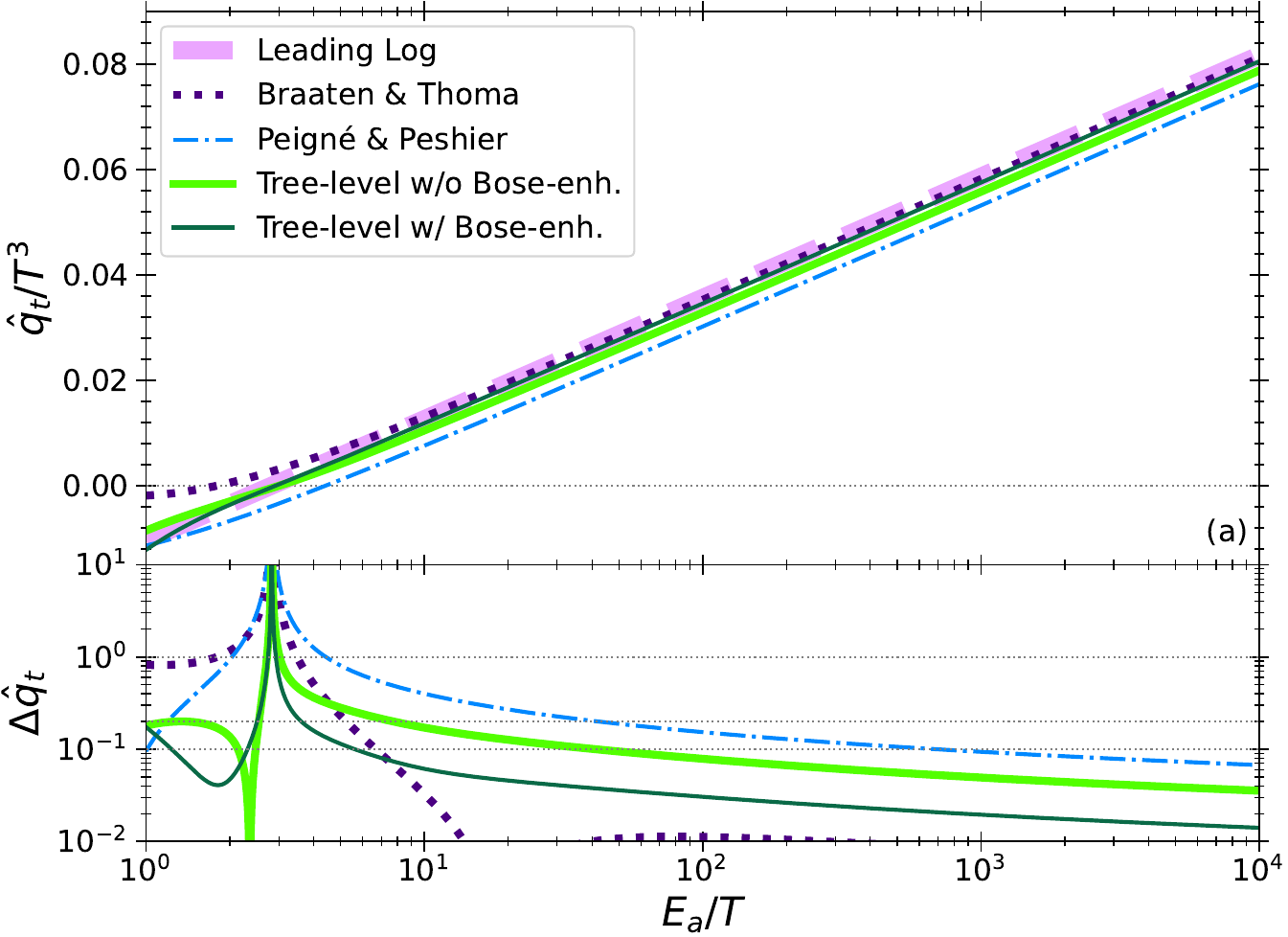}
    %\caption{ }
\end{subfigure}
\begin{subfigure}{\linewidth}
	\includegraphics[width=\linewidth]{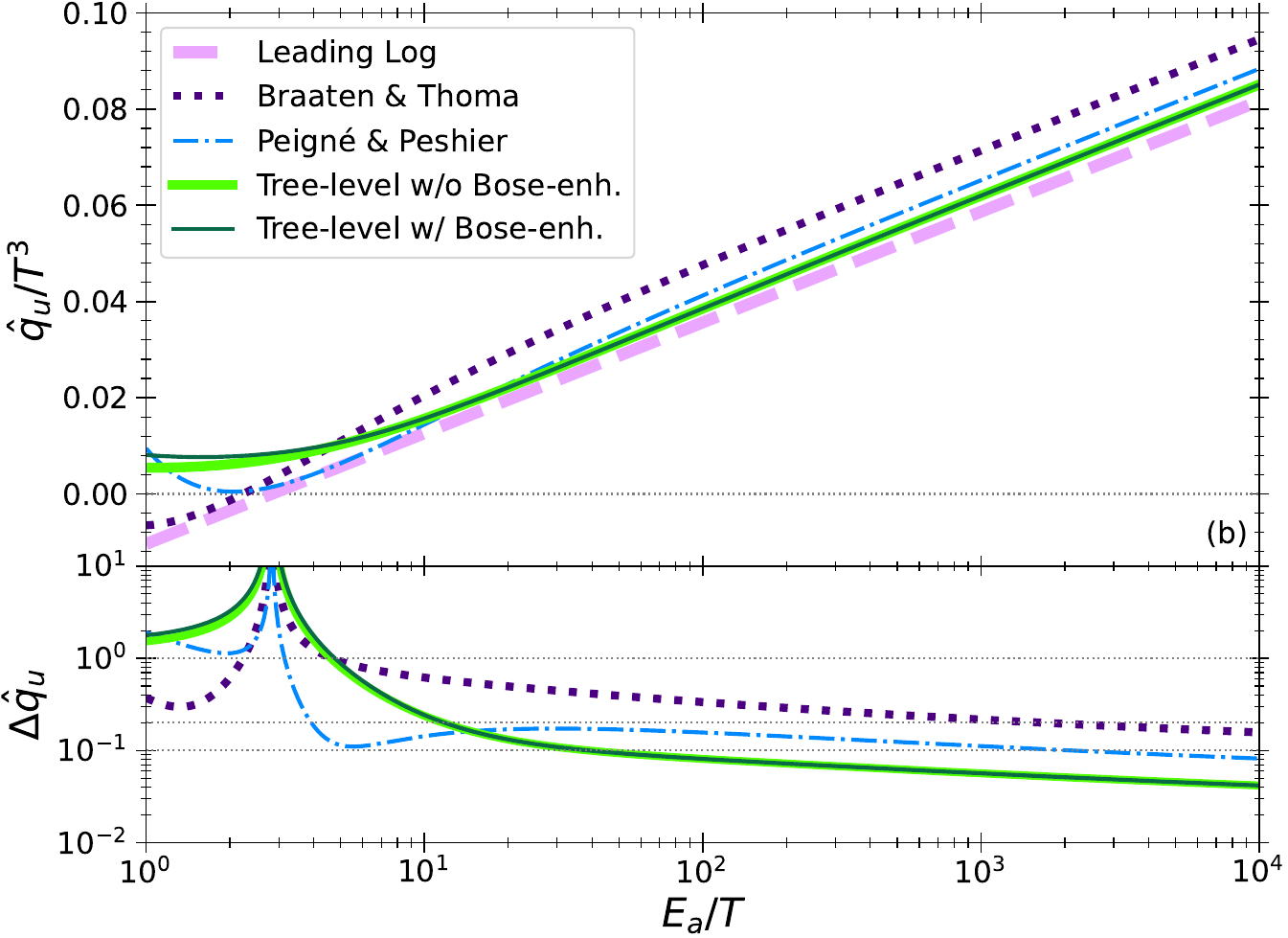}
    %\caption{$\langle q^2_\perp\rangle_L/L$ from the $u$-channel contribution of the $gg\to gg$ scattering matrix element.}
\end{subfigure}
\begin{subfigure}{\linewidth}
	\includegraphics[width=\linewidth]{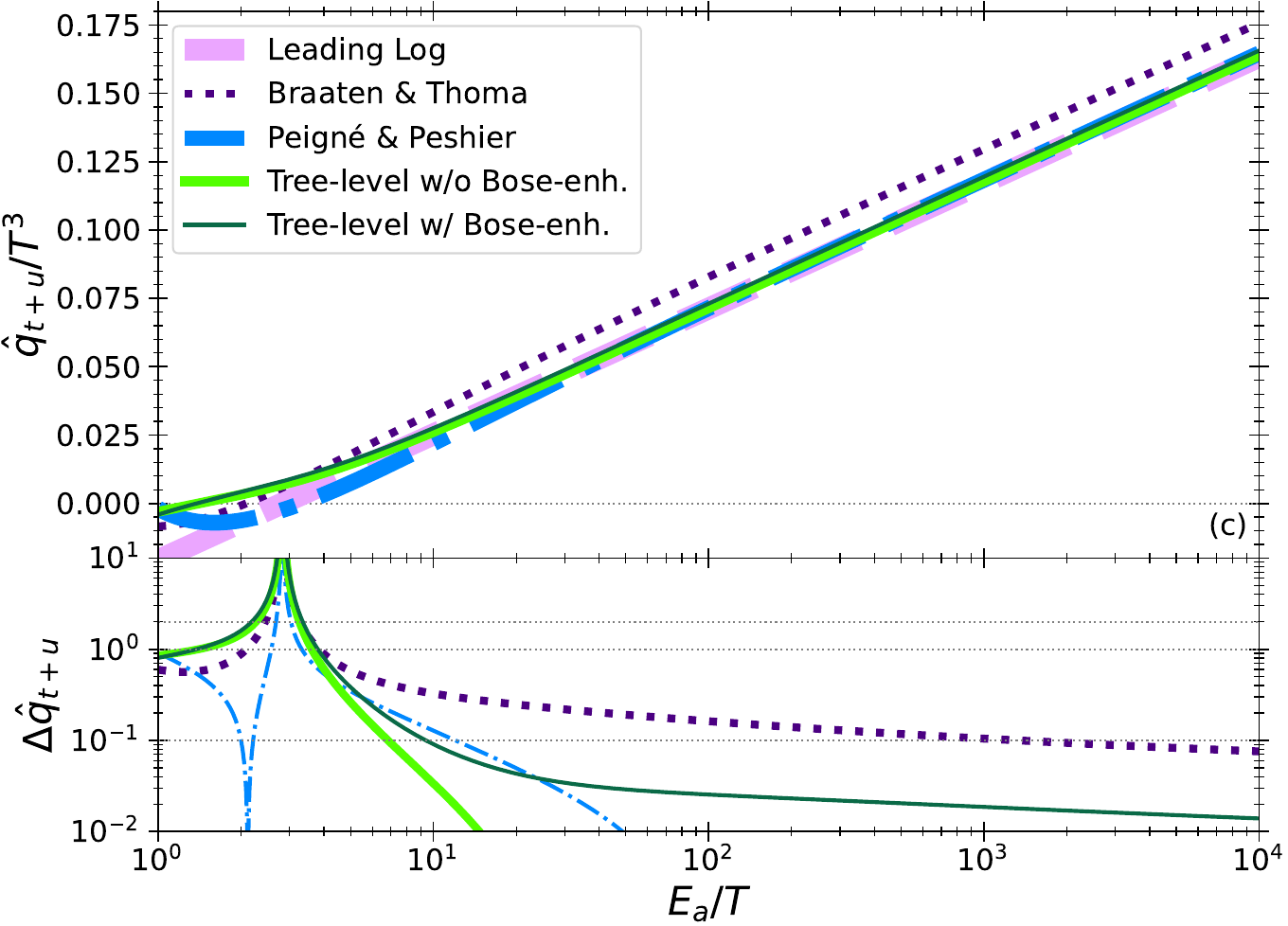}
    %\caption{$\langle q^2_\perp\rangle_L/L$ from the $t+u$-channel contributions of the $gg\to gg$ scattering matrix element.}
\end{subfigure}
    \caption{Transverse momentum broadening $\langle q^2_\perp\rangle_L/L$ was computed for $\alpha_s=0.3$ from various approaches. The $t$-channel, $u$-channel and $(t+u)$-channel contribution of the $gg\to gg$ scattering matrix element are presented in (a) through (c), respectively.}
\label{fig:q_hat}
\end{figure}

The $u$-channel contribution to $\langle q^2_\perp\rangle_L/L$ is obtained using Eq.~(\ref{eq:qhat_u}) and Eq.~(\ref{eq:qhat_u_PP}), representing the full and approximate result. Figure~\ref{fig:q_hat} depicts all approaches considered. A recent matching between the leading-log $2\to2$ scattering to HTL rates~\cite{Shen:2014nfa}, shows that the region where HTL resummations are sizeable starts at $x_a\sim 10$ and continues to grow as $x_a$ decreases, which one should keep in mind when examining the results in Fig.~\ref{fig:q_hat}. While the Peign\'e \& Peshier \cite{Peigne:2007sd,Peigne:2008nd} approach does include more of the kinematic phase space away from the ultraviolet behavior in the asymptotic regime considered by Braaten \& Thoma \cite{Braaten:1991jj,Braaten:1991we}, it still does not include enough of the kinematic phase space to truly cover tree-level $2\to2$ scattering processes. Once the full tree-level phase space is included along with Bose-enhancement, the final result for $\hat{q}$ fortuitously approaches Eq.~(\ref{eq:qhat_leading-log}), which neglects Bose-enhancement effects, as noted before. Indeed, such a happenstance between the leading-log and the full tree-level result did not occur for the $\frac{d^3 R_{t+u}}{d^3 p_a}$. If Bose-enhancement is included in Eq.~(\ref{eq:qhat_leading-log}), then the leading-log result increases by $\sim37\%$, clearly overestimating the true value: see dark green curves in Fig.~\ref{fig:q_hat}. As Bayesian model-to-data comparisons, such as Refs. \cite{JETSCAPE:2021ehl,JETSCAPE:2023ikg,JETSCAPE:2024cqe}, typically include a parameter that controls the overall normalization for $\hat q$, that $\sim37$\% enhancement is not a source of significant concern for the direct extraction of $\hat q$ itself. However, Bose-enhancement and detailed accounting of $2\to2$ kinematics does change the overall normalization and shape of the scattering rate, and thus the scattering probability, which, as discussed in the previous section, remains to be addressed in Monte Carlo simulations.

The recent constraint on $\hat q$ from the JETSCAPE Collaboration \cite{JETSCAPE:2024cqe} has the Bayesian software infrastructure needed to add $\Delta \hat q$ as a theoretical systematic uncertainty to any approximate parametrization of $\hat q$. As the energy dependence of $\hat{q}$ is an input to the Bayesian analysis of the JETSCAPE Collaboration \cite{JETSCAPE:2024cqe}, including $\Delta\hat q$ can be done without needing to simulate jets using Latin hypercube sampling, thus significantly reducing the computational resources needed to include $\Delta\hat q$ theoretical systematic uncertainty. Thus, depending on the model for $\hat q$, a {\it quantified} theoretical systematic uncertainty to be used in future Bayesian analysis is at hand. 

Unlike $\hat q$, the scattering rate's tight dependence onto the temperature profile probed by each parton traversing the QGP is such that theoretical uncertainty on the rate can be estimated by resampling, via Latin hypercube, the jet-medium Monte Carlo simulations. Another potential option is to use transfer learning methods~\cite{Paquet:2023rfd, Liyanage:2022byj} to account for the discrepancies between different approaches when obtaining the scattering rate. 

\textbf{\textit{The drag coefficient $\hat{e}$.}}--- The energy loss, or energy drag, transport coefficient $\hat e$ is defined as 
\begin{align}
\hat{e}&=\frac{1}{2E_a}\left[\prod_{i=b,1,2}\int\frac{d^3 p_i}{\left(2\pi\right)^3 2E_i}\right] \int \frac{d^4 q}{(2\pi)^3} f_b\left(p_b\right) \left[1 \pm f_2\left(p_2\right)\right]\nonumber\\
&\times(2\pi)^4\delta^{(4)}\left(p_a-p_1-q\right)\delta^{(4)}\left(p_b+q-p_2\right) \omega \, \overline{\left|\mathcal{M}_{a,b\to1,2}\right|^2}\nonumber\\
&=\frac{\langle\omega\rangle_L}{L},\label{eq:ehat_matrix}
\end{align}
where $q^\mu=(\omega,\vec q)$. This coefficient has been explored in the past by both Braaten \& Thoma as well as Peign\'e \& Peshier, though not for the $gg\to gg$ process. Although this coefficient has not been as extensively employed as $\hat q$ within current Bayesian analysis, e.g. \cite{JETSCAPE:2024cqe,JETSCAPE:2023ikg,JETSCAPE:2022jer,JETSCAPE:2021ehl}, such a Bayesian analysis should occur in the future in an effort to reduce the bias in Bayesian constraints on $\hat q$. For tree-level scatterings in the $q^2\approx -q^2_\perp$, $E_b\gg q_\perp$ and $E_a>E_b$ limit, one obtains \cite{long_paper_in_prep}
\begin{align}
\hat e_t\simeq \frac{9\alpha_s^2T^2}{4\pi^4}\ln\left(\frac{2x_a}{z^2}\right)\left[\frac{\pi^2}{6}\right] + O\left(e^{-x_a}\right),
\end{align}
with $\hat e_u=\hat e_t$. The Braaten and Thoma approach yields
\begin{align}
\hat e_t&\simeq\frac{9\alpha_s^2T^2}{4\pi^4}\left\{\left[\frac{\pi^2}{6}-\frac{2\zeta(3)}{x_a}\right]\ln\left(\frac{4x_a}{z^2}\right)-\frac{\gamma\pi^2}{6}+\zeta'(2)\right.\nonumber\\
&\left.\qquad\qquad+\frac{(2\gamma-1)\zeta(3)}{x_a}-\frac{2\zeta'(3)}{x_a}\right\}+O\left(\frac{z^2}{x^2_a}\right),\\
\hat{e}_u&\simeq\frac{9\alpha_s^2T^2}{4\pi^4}\left\{\frac{4\zeta(3)x_a}{z^2}+\left[\frac{2\zeta(3)}{x_a}-\frac{\pi^2}{6}\right]\ln\left(\frac{4x_a}{z^2}\right)\right.\nonumber\\
&\left.\qquad\qquad\quad-\frac{2\pi^4}{15z^2}+\frac{(\gamma-1)\pi^2}{6}-\zeta'(2)+\frac{2\zeta'(3)}{x_a}\right.\nonumber\\
&\left.\qquad\qquad\quad+\frac{(3-2\gamma)\zeta(3)}{x_a}\right\}+O\left(\frac{z^2}{x^2_a}\right),
\end{align}
which changes the functional dependence in the ultraviolet regime by changing some coefficients, providing additional constants, as well as $x^{-1}_a$ terms. 
\begin{figure}[H]
\centering
\begin{subfigure}{\linewidth}
	\includegraphics[width=\linewidth]{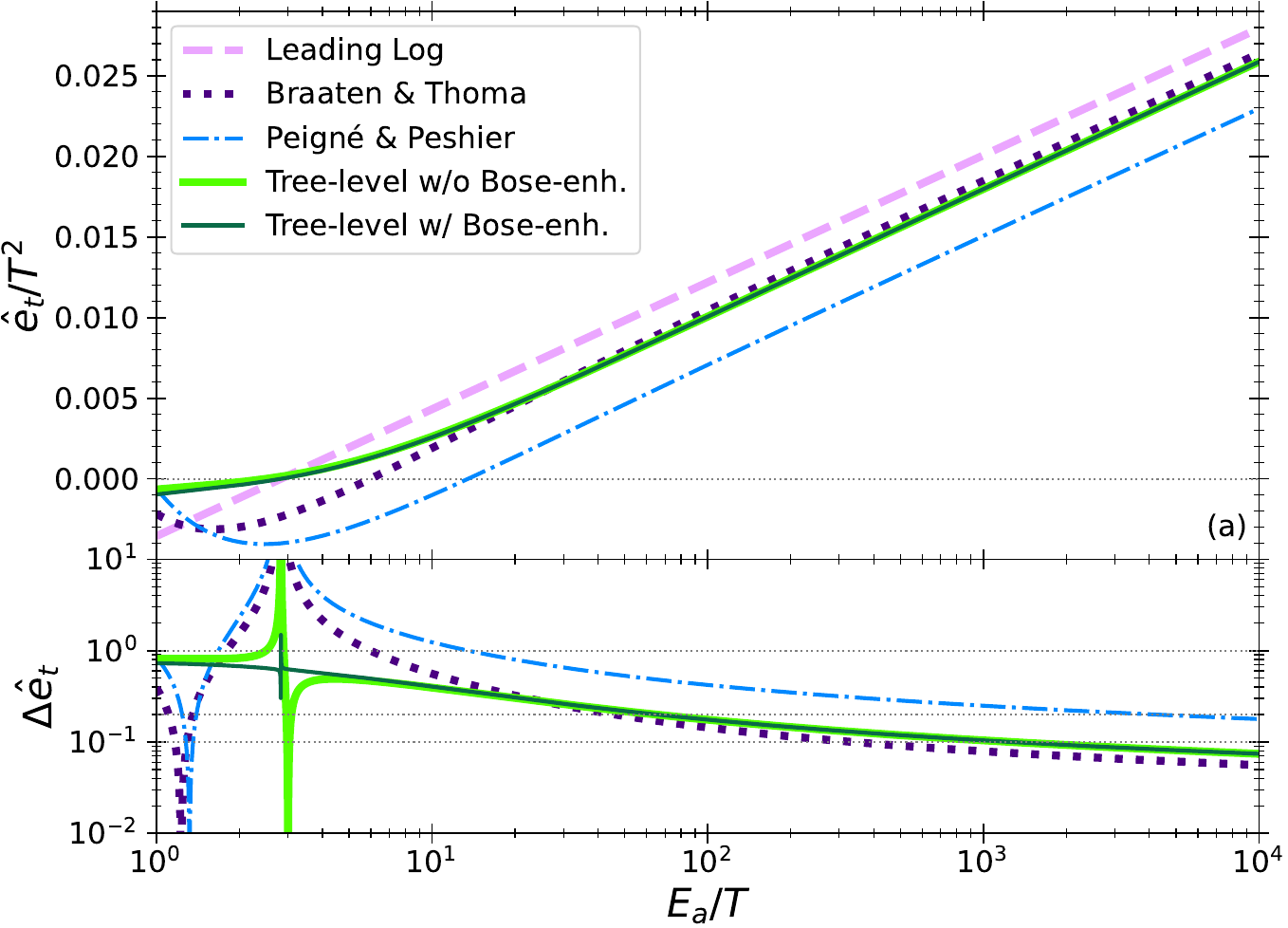}
    %\caption{ }
\end{subfigure}
\begin{subfigure}{\linewidth}
	\includegraphics[width=\linewidth]{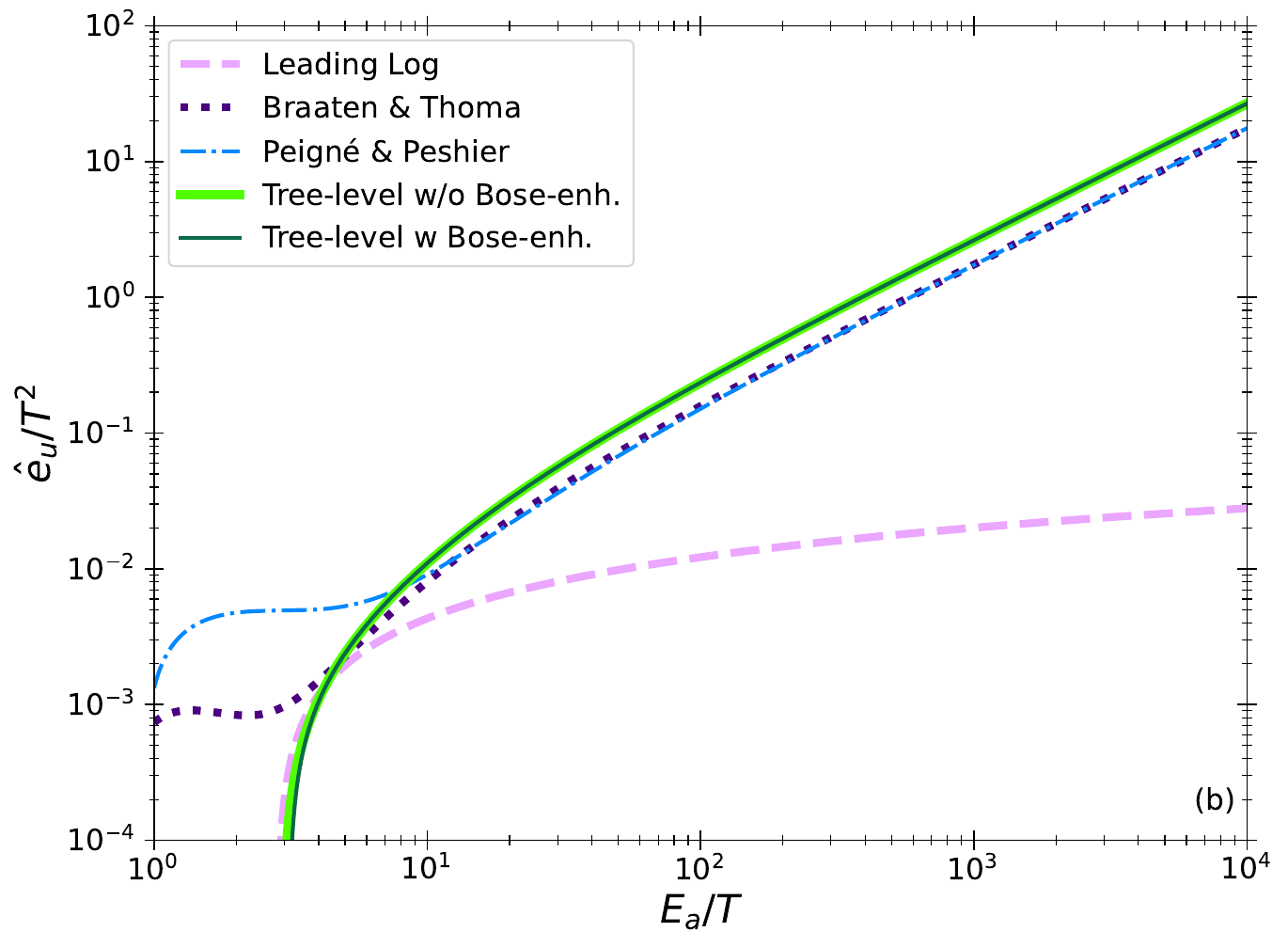}
    %\caption{$\langle q^2_\perp\rangle_L/L$ from the $u$-channel contribution of the $gg\to gg$ scattering matrix element.}
\end{subfigure}
\begin{subfigure}{\linewidth}
	\includegraphics[width=\linewidth]{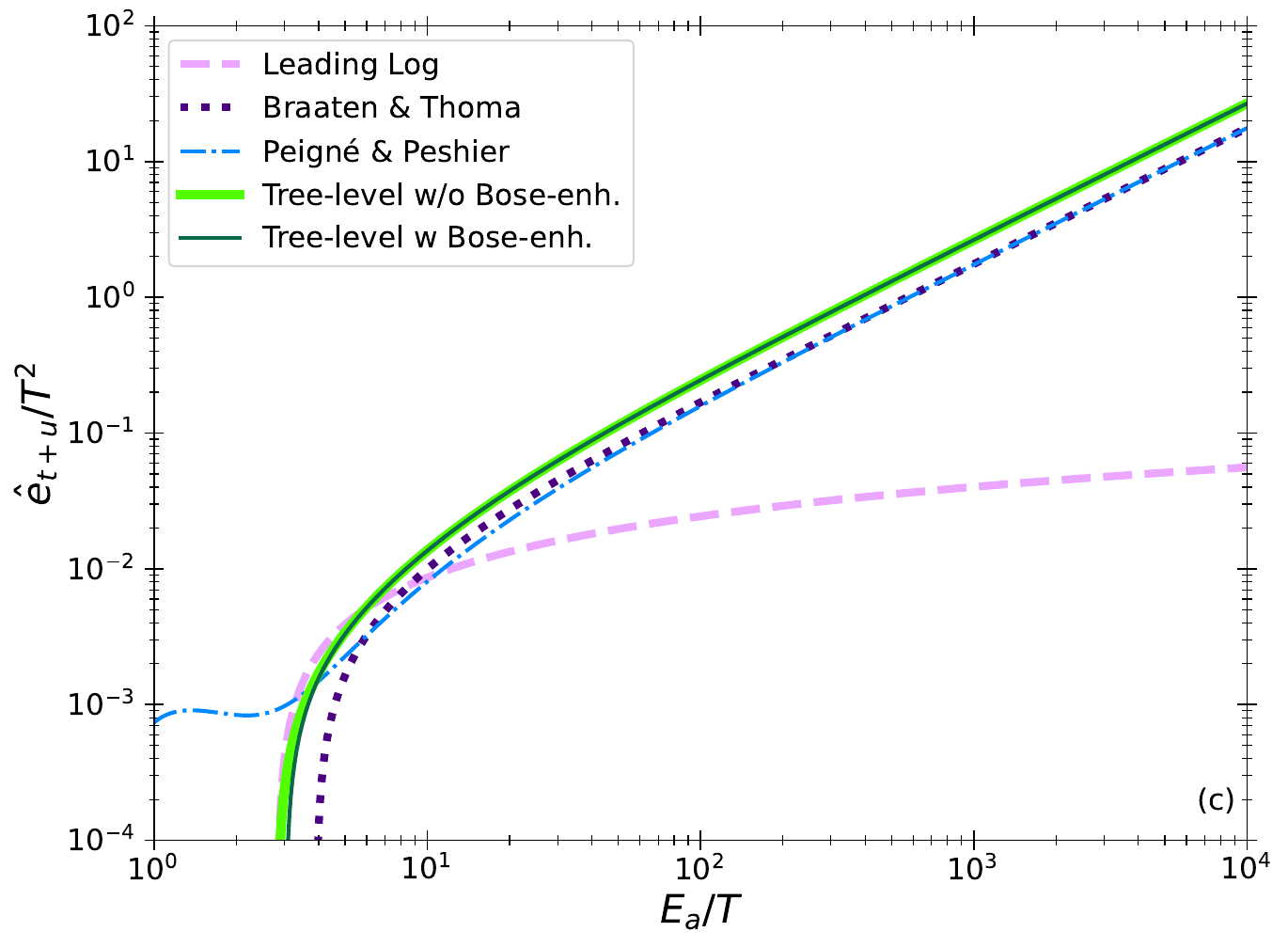}
    %\caption{$\langle q^2_\perp\rangle_L/L$ from the $t+u$-channel contributions of the $gg\to gg$ scattering matrix element.}
\end{subfigure}
    \caption{Energy loss $\langle \omega\rangle_L/L$ was computed for $\alpha_s=0.3$ from various approaches. The $t$-channel, $u$-channel and $(t+u)$-channel contribution of the $gg\to gg$ scattering matrix element are presented in (a) through (c), respectively.}
\label{fig:e_hat}
\end{figure}

The Peign\'e \& Peshier result keeps the functional form the same, and corrects coefficients of the Braaten \& Thoma approach, thus yielding
\begin{align}
\hat e_t&\simeq\frac{9\alpha_s^2T^2}{4\pi^4}\left\{\left[\frac{\pi^2}{6}-\frac{2\zeta(3)}{x_a}\right]\ln\left(\frac{4x_a}{z^2}\right)-\frac{(1+\gamma)\pi^2}{6}\right.\nonumber\\
&\left.+\zeta'(2)+\frac{(1+2\gamma)\zeta(3)}{x_a}-\frac{2\zeta'(3)}{x_a}\right\}+O\left(\frac{z^2}{x_a^2}\right),\label{eq:ehat_t_PP}\\
\hat{e}_u&\simeq\frac{9\alpha_s^2T^2}{4\pi^4}\left\{\frac{4\zeta(3)x_a}{z^2}+\left[\frac{4\zeta(3)}{x_a}-\frac{\pi^2}{3}\right]\ln\left(\frac{4x_a}{z^2}\right)-\frac{2\pi^4}{15z^2}\right.\nonumber\\
&\left.+\frac{\gamma\pi^2}{3}-2\zeta'(2)+\frac{4\zeta'(3)}{x_a}+\frac{2(1-2\gamma)\zeta(3)}{x_a}\right\}+O\left(\frac{z^2}{x_a^2}\right).\label{eq:ehat_u_PP}
\end{align}

Going beyond Peign\'e and Peshier, the tree-level result is
\begin{align}
\hat{e}_t&\simeq\frac{9\alpha_s^2T^2}{4\pi^4}\left[\left(\frac{\pi^2}{6}-\frac{2\zeta(3)}{x_a}\right)\arcsinh\left(\frac{x_a}{z}\right)+B^-_1(z)\right.\nonumber\\
&\left.-\frac{\pi^2\sqrt{x_a^2+z^2}}{3x_a}-\frac{B^-_2(z)}{x_a}+\frac{2A^-_1(z)}{x_a}\right]+O\left(e^{-x_2}\right),\label{eq:ehat_t}\\
\hat{e}_u&\simeq\frac{9\alpha_s^2T^2}{4\pi^4}\left[\left(\frac{2\zeta(3)}{z^2}-\frac{\pi^4}{15x_az^2}+\frac{\pi^2}{3x_a}\right)\sqrt{x_a^2+z^2}\right.\nonumber\\
&+\frac{A^-_1(z)x_a}{z^2}+\left(\frac{4\zeta(3)}{x_a}-\frac{\pi^2}{3}\right)\arcsinh\left(\frac{x_a}{z}\right)\nonumber\\
&\left.-\frac{A^-_2(z)}{z^2}-2B^-_1(z)-\frac{2A^-_1(z)}{x_1}+\frac{2B^-_2(z)}{x_a}\right]+O\left(e^{-x_2}\right),\label{eq:ehat_u}
\end{align}
if Bose-enhancement effects are neglected. A similar change in the functional dependence as that seen in Eq.~(\ref{eq:rate}) is observed again here. As a result of a careful calculation, one notices that $\hat e_t$ in Eq.~(\ref{eq:ehat_t}) now approaches the Braaten \& Thoma result, more so than the Peign\'e \& Peshier one, as depicted in Fig.~\ref{fig:e_hat}(a). Note that $\Delta \hat e$ is defined analogously to Eq.~(\ref{eq:Delta_qhat}). Inspecting $\Delta \hat e_t$ in Fig.~\ref{fig:e_hat}, one notices that the $x_a$ dependence in $\hat e_t$ of the non-Bose-enhanced result follows the Bose-enhanced result closely. Thus, to a very good approximation, the Bose-enhanced $\hat e_t$ is analytically given by Eq.~(\ref{eq:ehat_t}).

Examining the $u$-channel contribution to $\hat e$ reveals that approximations akin to those in Eq.~(\ref{eq:approx}), when applied to $A^-_i(z)$ and $B^-_i(z)$, affect all approximate expressions for $\hat e$ obtained using methods devised by previous authors.

While the leading-log expression for $\hat q$ behaves similarly to other approaches, the leading-log expression of $\hat e$ in the $u$-channel leads to an incorrect $x_a$ dependence, giving a clear limitation of the leading-log approximation.

\textbf{\textit{Discussion and outlook}}--- The mathematical details for the results presented herein are available in Ref.~\cite{long_paper_in_prep}. Having established the theoretical approach to calculate scattering rates and jet-medium transport coefficients in the ultraviolet regime, revising all calculations involving $2\to2$ matrix elements, and constructing estimates of theoretical systematic certainties as shown here, is to be explored next. Focus will be given towards novel jet-medium transport coefficients relevant for the higher twist calculations in Refs.~\cite{Kumar:2025egh,Kumar:2025asj}, specifically those involving Glauber quarks. Beyond this, improving the jet-medium Bayesian analysis by including Bayesian model discrepancy \cite{Jaiswal:2025hyp,Jaiswal:2025deb} to account for theoretical systematic uncertainty in areas where estimating theory uncertainty is challenging, such as hadronization mechanisms, is a key area of future improvement of jet-medium Bayesian analysis. A crucial component to addressing this uncertainty also involves a simultaneous revision of hadronization models themselves. The JETSCAPE Collaboration has been working in this direction using hybrid hadronization \cite{JETSCAPE:2025wjn}, while additional research is ongoing regarding quark recombination/coalescence \cite{Fries:2025jfi,Fries:2025brq}. Thus, combining the Bayesian model discrepancy with enhanced recombination/coalescence models, for situations where a first principles calculations cannot provide an estimate of the theoretical systematic uncertainty, along with a careful accounting of partonic level uncertainties where theory does provide an estimate for uncertainties, is a critical milestone for future Bayesian analysis that will improve the reliability of jet-medium transport coefficient constraints. The ongoing pursuit of phenomenological constraints of relativistic heavy-ion collision simulations requires accounting for theoretical and experimental uncertainties with increasing precision to ensure continued success.

\textbf{\textit{Acknowledgments.}}--- The authors thank Charles Gale for valuable discussions and comments on this work. This work was supported by the Canada Research Chair under Grant Number CRC-2022-00146, the Natural Sciences and Engineering Research Council (NSERC) of Canada under Grant Number SAPIN-2023-00029, and the Canadian Foundation for Innovation John R. Evans Leaders Fund under Grant Number 44100.

\textbf{\textit{Data availability.}}--- No data were created or analyzed in this study.

\bibliography{references}
\end{document}